\documentclass{ws-rv10x7}
\usepackage{ws-rv-van}     
\usepackage{ws-rv-thm}     
\usepackage{subfigure}
\makeindex
\begin{document}

\chapter[The Random First Order Transition Theory of Glasses in Real Space-Time]{The Random First Order Transition Theory of Glasses in Real Space-Time: Instantons, Strings, Flames and Flows\label{ch1}}

\author[P.G. Wolynes]{Peter G. Wolynes\footnote{pwolynes@rice.edu}\footnote{This
manuscript will appear in the book ``Spin Glass Theory and Far Beyond -
Replica Symmetry Breaking after 40 Years''.}}

\address{Rice University, Departments of Chemistry and Physics,\\ and Center for Theoretical Biological Physics,\\
6100 Main Street, Houston, Texas 77005, USA}

\begin{abstract}
Understanding structural glasses using the Random First Order Transition theory requires a description in both real space and in time, taking into account sample history. A variety of nonlinear objects enter this description, in field theory terms, when there is replica symmetry breaking: instantons, strings, flames and flows. We focus on this real space description and use it to describe the crossover to the energy landscape regime, dynamical heterogeneity, aging and the formation of shear bands in glasses under stress.
\end{abstract}

\body

\section{Introduction}\label{sec1}
Water, Air, Earth, and Fire. Many ancient thinkers recognized these as the elementary aspects of the Natural World, explaining the phenomena they observed as involving their interchange. Arguably, much of modern theoretical physics, in using field theories, adopts the same stance, albeit employing thousands of years of increasing mathematical sophistication. Other ancient thinkers conceived of natural phenomena as universally arising from the motions of ultimately small constituents, atoms. Reconciling these two views comprises a big part of the business of modern statistical physics.

The ancients’ view of the relationship between Air and Water involved their interchange through the processes of rarefaction and densification.\cite{schrodinger1996nature} Van der Waals’s theory of the continuity of the gas and liquid phases provided a quantitative basis for this old notion.\cite{van2004continuity} The density, a single number, then distinguishes the two forms through a scalar field which may then vary in space. The fluctuations of this field in space provide the mechanism of interconversion through nucleation of the gas to the liquid form through forming droplets of the high-density form within the lower density form\cite{zeldovich2014selected} and the reverse process of forming low density bubbles in the high-density form. Strikingly, field fluctuations also provide the explanation of the singular mathematical variations of many physical properties when these two continuously related forms become very similar in density – near the so-called critical point.\cite{wilson1983renormalization} 

The ancients were more puzzled by the relationship of Earth to Air and Water which differed merely by density. Atomists suggested the essential connection had to do with the shapes of the atoms, arguing that some shapes of atoms, when put together at sufficiently high density, would ineluctably form a rigid body. In more modern times, Kepler articulated a more sophisticated view in his monograph \textit{De Nive Sexangula}.\cite{kepler1966strena} He suggested that the atoms themselves need not have complex shapes, but that even spheres would arrange themselves (like piles of cannonballs) in a highly symmetric fashion (what we now call a face-centered cubic close packing). Rearranging this nearly unique structure would require elaborate motions of many atoms to retain the symmetry, so a snowflake must move as a rigid object, in contrast to what happens for gases or a liquid, which can flow in complex ways. The symmetry of the packing, thus, not only explained the six-sidedness of snowflakes, but also owing to the structural uniqueness, the solid form’s rigidity. Twentieth century physicists would call this situation a “broken symmetry” because the six-sided shape of a snowflake is not as symmetrical as a sphere, the usual shape of a liquid drop. Anderson, Landau, and de Gennes have emphasized “broken symmetry” as the origin of rigidity throughout the zoo of forms of condensed matter, superconductors, liquid crystals, and  magnets.\cite{anderson1997concepts,de1992soft,ginzburg2004nobel}  Members of the condensed matter zoo then are described through the interactions of several fields (each one possibly more complex than a scalar), allowing one to still use the concept and its powerful field theory mathematics. A puzzle remains: macroscopically, not all matter that appears to be rigid shows the faceting that is characteristic of the snowflake. There are solid spherical glass beads, as well as, still more elaborately shaped glassy materials that are rigid. One partial resolution of this puzzle led to the deep ideas of modern metallurgy: real solids, even if crystalline, are not usually single crystals, but instead are jumbles of small crystallites (each crystallite containing still very many atoms, making them act rigidly), which contain spatially localized defects that do allow such solids to be deformed, albeit with greater difficulty than a liquid or a glass, allowing them to be made into a variety of useful shapes. The resulting beautiful theory of grain boundaries, disclinations, and dislocations\cite{taylor1934mechanism,Polanyi1934,orowan1940problems,read1950dislocation} thus preserves the idea of broken symmetry as the origin of rigidity, but provides mechanisms for structural variety and interconversion of macroscopic forms. The defects of real crystalline solids represent atomic level near singularities in otherwise smooth fields describing the symmetries. Glasses challenge this harmonious paradigm of traditional metallurgy and solid state physics: at the shortest length scales, to the extent that we can probe them, the structures of glasses are so varied that it is very hard to make an assignment of a few broken symmetry fields to characterize their structures. This is not to say that making such symmetry assignments is completely impossible: many ideas for locally broken symmetries in glasses such as the presence of icosahedral order have been put forward. These notions have enjoyed varying degrees of success.\cite{steinhardt1981icosahedral,sachdev1985order} Also, recently, people have been trying to use the tools of artificial intelligence and machine learning to try to identify local order parameters in simulated liquids and glasses. It appears the number of broken symmetries then must be large.

The Random First Order Transition Theory (RFOT) of glasses avoids trying to force the description of amorphous structure into the straightjacket of traditional field theory, which uses only a small number of fields. Instead, the RFOT theory starts from the assumption that there is a statistically large number of possible arrangements of even the simplest (spherical) atoms that are, nonetheless, mechanically stable. Each of these structures can be effectively rigid since they represent local minima of the energy. This idea goes back at least to Descartes.\cite{descartes} It was made most concrete in modern times by J. D. Bernal\cite{bernal1964bakerian} who showed such “random” structures can explain x-ray diffraction from liquids. In the context of the magnetism of alloys, Edwards and Anderson suggested a statistically large number of stable magnetic patterns was possible for spins that have quenched random interactions. Many models with quenched random interactions,\cite{edwards1975theory,edwards1976theory} for example, the p-spin glasses and the Potts spin glasses\cite{gross1985mean,kirkpatrick1987dynamics} also have many metastable minima. More recently, the existence of statistically large numbers of fields has emerged self-consistently in systems without quenched disorder in studies of hard spheres assemblies in very high dimension.\cite{1987Kirkpatrick:1987ek,parisi2020theory} In these mean field theories, which are based on the statistics of the patterns one finds the emergence of a multiplicity of metastable states. Each one of these states is rigid, and could last forever in the absence of thermal motion, like a cold amorphous snowflake. This phenomenon is called Replica Symmetry Breaking.\cite{mezard1987spin} The first decades of exploring Replica Symmetry Breaking in the magnetic alloys culminated in the exact solution of the Sherrington-Kirkpatrick spin glass model with infinite range interactions by Parisi.\cite{parisi1979infinite} It took some time for the mathematical apparatus of this exact solution to be understood in physical terms,\cite{mezard1987spin} but this analysis provides a firm place to start to study real glasses. Yet, one must go further.

Like the theory of real crystalline solids and the theory of the liquid-gas transformation, understanding laboratory glasses requires going beyond the mean field theory, which treats problems of stability globally. One must describe the local motions of the atoms in real space and time to achieve a description of glasses, as we find them, in finite dimensions. Here, we will describe the dynamics of glasses and glass forming systems using extensions of replica symmetry breaking ideas that allow the statistics of the atomic arrangements to vary in space and time and to have locally near singular behavior: in the field theory language, these variations are described using the ideas of instantons, strings, flames, and flows. While far from mathematically rigorous, this development of replica symmetry breaking that is inhomogeneous in space and time provides a reasonably simple framework that explains a wide range of phenomena seen in glasses. It also explains several quantitative regularities that connect the thermodynamics and the kinetics of processes in the liquid state as the liquid transforms into a rigid glassy state.\cite{lubchenko2015theory,2007Lubchenko:2007cca} The limitations of space in this chapter forces us to be telegraphic both as to the mathematical details of this complex of theoretical developments (for which we refer the reader to a more complete yet still pedagogical review\cite{lubchenko2015theory}) as well as to displaying the more detailed comparisons of the dynamical theory with experiments (for which, again, a recent review exists\cite{ediger2021glass}). Since the theory of these local motions is clearly approximate and has “loose ends”, we note that it is the harmony it has with experiment that we find convincing evidence for the basic ideas.

\section{Self-Generated Randomness and the Local Landscape}
Aperiodic energy minima, reflecting locally stable structures, have been generated in several ways. Bernal literally jammed hard spherical balls into a balloon and by sucking the air out, compressed the packed balloon into a high-density conglomeration that was sensibly rigid.\cite{bernal1964bakerian} By deconstructing by hand the resulting assembly, he was able to characterize many statistical aspects of such a locally stable, albeit, finite structure, including the statistics of the number of close contacts and the extent of local icosahedral order. As an alternative approach, he built similar jammed structures by hand, by randomly adding hard spheres individually to a seed structure. In this way, he constructed an aperiodic, but finite large cluster. He then carried out analogous procedures using a digital computer in one of the first uses of computers in statistical mechanics. At IBM, Bennett extended this computer assembly approach.\cite{bennett1972serially} In a more systematic fashion, Stillinger and Weber used computational quenching by steepest descent on the potential energy surface to create aperiodic structures that are minima of the potential energy. Each one of the minima generated in this way is near to a simulated liquid configuration that was sampled in an equilibrated molecular dynamics run.\cite{stillinger1982hidden} By doing this, the energy landscape can be fully tiled with basins of attraction. The assemblies generated in all these various ways seem to be as mechanically stable as Kepler’s periodically stacked cannon balls were. These studies thus established that aperiodic “crystallites” of quite large size are stable in the absence of thermal motion. One could question whether such procedures could be extended into the thermodynamic limit, but as a starting point to understanding the stability of glasses in the laboratory, achieving a truly infinite aperiodic crystal is unnecessary, just as polycrystalline metals are sensibly solid even though their ordered parts are not strictly infinite, because they, nevertheless, contain many atoms, which all must move in concert.

The local stability of these constructed aperiodic assemblies to thermal fluctuations can be assessed using the same tools that have been used to study the local stability of periodic crystals (e.g. self-consistent phonon theory\cite{1984Stoessel:uz} and density functional theory\cite{1985Singh:1985dc}) in the context of freezing into periodic structures. In this way, one establishes that there are not only energy minima, but also that there are many local minima of a free energy functional each of which corresponds to an aperiodic density pattern. In other words, given a free energy functional $\mathcal{F}[\{\rho(\underline{r})\}]$, there are numerous local minima that are aperiodic density waves satisfying the functional derivative condition $\frac{\delta \mathcal{F}}{\delta \rho} = 0$. 

The existence of local free energy minima that are not periodic does not guarantee that any such structure can ever be a global minimum of the free energy -- almost certainly, for molecules with the simplest shapes, instead, periodic density waves corresponding to the stable crystals will be lower in free energy. According to thermodynamics and ergodicity, such systems, while being aperiodic, will ultimately crystallize, so the observed phenomena of glasses must involve only metastable free energy minima.\cite{1985Solids:wd} Furthermore, in general, a single individual aperiodic density pattern cannot be more stable than the entire set of all such aperiodic patterns, which are thermodynamically extensive in number, unless some sort of an ideal glass transition is reached at very low energy. A single aperiodic structure always thermodynamically resembles a superheated “crystal”, in that each aperiodic crystal is only metastable with respect to the liquid, which corresponds with an average of such structures, having a uniform density $\rho(\underline{r}) = \bar{\rho}$. The dynamics in the vicinity of a given local free energy minimum can thus be monitored using a local scalar field that measures the similarity of a given aperiodic density pattern to the density wave pattern that describes the closest local minimum. In the theory of spin glasses, this quantity is called an overlap order parameter. The corresponding quantity for a structural glass can actually be experimentally monitored globally through the time delayed structure function $\langle \delta \rho(\underline{r},t)\delta \rho(\underline{r}',t+\tau)\rangle$. In principle, the local value of this order parameter would be accessible to observation with appropriate microscopes while its global average can be easily found via neutron scattering. In deeply supercooled liquids, the time delayed structure function has a long plateau for large, but intermediate values of the delay time, $\tau$. The value of this plateau level plays the role of the Edwards-Anderson order parameter in spin glasses, which are systems that have quenched disorder in the interactions, for which the plateau can last forever. Of course, the structure function ultimately decays to zero through a process of escape from the starting metastable local free energy minimum to some other ones, which leads to the end of the plateau for large enough time. For practical people, the plateau is long enough to be quite well defined, but obviously, the final decay with $\tau$ presents difficulties for rigorous mathematical people, because of the awkward interchange of the limits of infinite plateau time and the thermodynamic limit.

\begin{figure}
\centerline{\includegraphics[width=8cm]{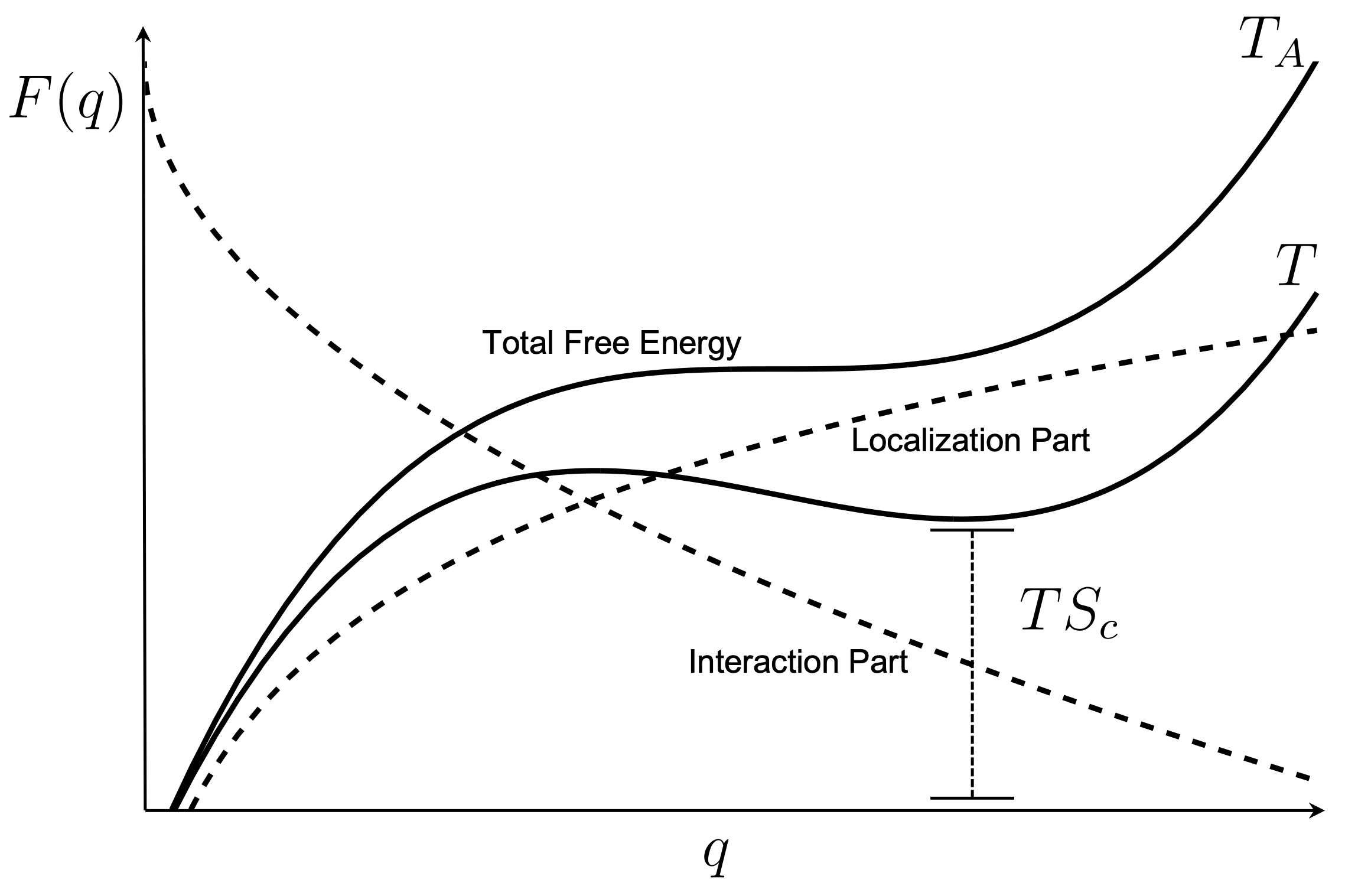}}
\caption{Schematic diagram of the free energy as a function of the structural overlap parameter $q$. There are two contributions: one coming from the entropy cost of localizing particles aperiodic lattice, the other coming from the stabilization arising from particles avoiding each other on that lattice. At the temperature $T_A$, a metastable minimum begins to appear which becomes more stable upon cooling. It is still less stable than the uniform system, owing to configurational entropy of the uniform liquid state. }
\label{fig:FreeEnergy}
\end{figure}

The free energy as a function of the plateau value $q$ if it were assumed to be uniform is sketched in Figure~\ref{fig:FreeEnergy}.
This free energy profile computed for a uniform value of the overlap order parameter resembles what we would find for an ordinary first order transition like the formation of periodic density wave in a crystal or a glass emerging from a liquid. Essentially, at this level, the initial comparison configuration, or more precisely, the density wave built on its nearest potential energy minimum, acts like what we would get for any other lattice, even though the initial pattern of atoms is not periodic in space. At high temperatures, there will be a large entropic cost for localizing the system near to this initial state. This cost is, however, partially overcome by the fact that the initial density wave pattern being relatively stable, avoids repulsive conflicts between molecules, so that at a sufficiently low temperature (or high enough density for hard spheres), the gain of stability from having localized the atoms, is prevented from making excursions into high energy steric conflict, thus allowing the aperiodic crystal to be, at least, metastable. Escaping this metastable minimum then will require activated dynamics, just as a superheated liquid needs to boil. If all the atoms had to disorder at once, the free energy barrier to escape would be extensive in $N$. The temperature at which metastability appears and where a secondary metastable minimum appears at finite $q$ is called $T_A$.

Above $T_A$, dynamics of the glassy liquid is collisional and can be described by mode coupling theory, but below this temperature mean field, mode coupling arguments, which are perturbative,  would lead to a finite value of $q$, just as the free energy analysis does. Mode coupling theory and density functional approaches are two sides of the same coin.\cite{wolynes1988aperioidic}

$T_A$ represents a spinodal temperature for stabilizing a metastable basin. On further cooling, the entropic advantage continues further to overcome the localization entropy cost. At some point, the two minima on the free energy profile could actually come close to coinciding in free energy. At this point, the entropic cost confining the system to being near a single configuration would not lose out, in free energy terms, to the multiplicity of static structures available at $q=0$. For this to happen, the logarithm of the number of alternate density patterns must no longer be extensive in system size. If such a point were to be reached, a given metastable minimum would remain stable forever. It would be an ``ideal glass''. At this level of description, the liquid-glass transition is hardly different from van der Waals's description of the liquid-gas transition or the theory of freezing into a periodic crystal. The transition temperature, where there would be no extensive configurational entropy advantage to being able to sample other structures, is called the Kauzmann temperature, $T_K$.

Such an impending entropy crisis was discussed by Kauzmann in his famous review.\cite{kauzmann1948nature} He inferred the possibility of such an impending entropy crisis based on experimental measurements of the configurational entropy of liquids that were obtained directly by calorimetry. Because of the necessity to extrapolate these data to very low temperatures, the actual existence of this ideal glass transition still enjoys some controversy. This concern, however, does not in any way, however, undermine RFOT arguments, which depend only on the existence of many metastable glassy states, not their global stability.

At this level, the final escape from the structure function plateau in which the system has been trapped near a metastable configuration proceeds exactly like the melting of a superheated crystal. Access to the other possible arrangements of atoms occurs by nucleating these alternate arrangements at specific points in space. A small region will cooperatively rearrange, so as to resemble one of the other global metastable structures. Because this motion takes place only locally, the free energy barrier for rearrangement in glasses is not extensive. The nucleation picture of metastable state escape we have just described has puzzled many. They wonder, “When a superheated liquid boils, each bubble continues to grow after it forms. Does this happen here?” Yes, this would happen, but because the new configuration that would be formed is typically just as stable as the starting one, there is a frictional brake on such further growth, because growth requires still more reconfiguration events to take place, albeit, at first, a bit faster perhaps. This frictional brake, modifies the transition state theory rate by a frictional Kramers correction.\cite{1988Onuchic:vnb} So, we see the process of nucleation must start over again to allow a macroscopic flow of the liquid. In any case, at the very same time, most of the neighboring regions to a given nucleation event also will have had a chance to make similar rearrangements. So, to a first approximation, yes, the rest of the glass will have already reconfigured by itself anyway, therefore, not requiring bubble growth from a single starting point.

Within the RFOT picture, glassy dynamics, then occurs by localized excitations or droplets. If the plateau value $q$ or overlap, which in mean field theory is a global order parameter, is ``promoted'' to being a time and space dependent field, this droplet corresponds to a saddle point of a field theoretical free energy functional
\begin{equation}
    \mathcal{F}[\{q(\underline{r})\}] =  \int d^{3} \underline{r}\left(F(q)+\frac{\kappa}{2}[\nabla q(\underline{r})]^{2}\right)
\end{equation}
if we assume $q$ varies only slowly in space. This saddle point solution is described by a nonlinear partial differential equation 
\begin{equation}
    \kappa \nabla^2 q = -\frac{\partial F}{\partial q}.
\end{equation}
The square gradient term in the functional represents the free energy cost of allowing $q$ to vary in space. The inhomogeneity costs reflect the fact that the intermolecular forces in real glasses are fairly short range. An instanton solution of the partial differential equation\cite{1987Kirkpatrick:1987cb,2005Dzero:tp,2009Dzero:tr} then will be made up of a central core where $q$ is nearly equal to 0. In this core region, the system accesses a set of possible states to which the initial configuration has very significantly rearranged. This core region can be called a local library of states.\cite{lubchenko2004theory,bouchaud2004adam} There will also be an outer region that has not yet transformed. This region will still have a large $q$ for a time, since the original structure started out globally rigid. Between the core and its surroundings, $q$ changes, allowing vestiges of the initial configuration to remain, but where a variety of similar states are being accessed. This effect is called ``wetting''. Near the spinodal $T_A$, this intermediate region could be quite large spatially. Again, using the analogy to the ordinary first order transition, Kirkpatrick and Wolynes\cite{1987Kirkpatrick:1987cb} suggested that the interface width would scale as $(T-T_A)^{-1/4}$. Of course, at the dynamical transition $T_A$ itself, this length diverges, but the barrier also vanishes at that point, so these droplets or instantons do not cost much free energy. Because of this, the overlap of one instanton with other instantons is inevitable, destroying the divergence of the correlation length predicted by the mean field calculation. While spinodals in mean field theory resemble critical points, in fact, spinodals (anywhere except at a critical point!) are crossovers when the interactions have finite range in three dimensions. 

\begin{figure}
\centerline{\includegraphics[width=5.2cm]{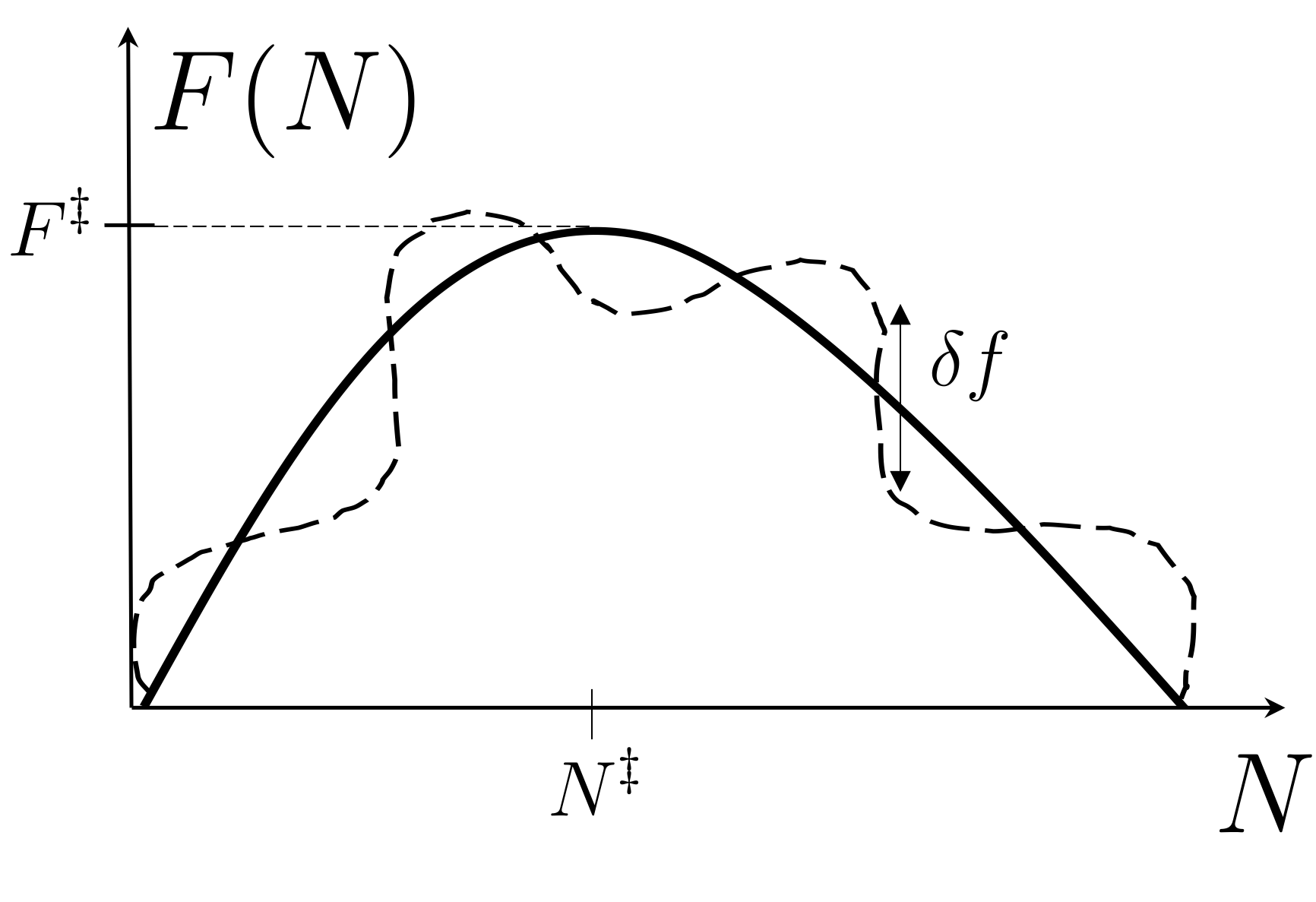}}
\caption{Sketch of the free energy profile for rearranging a region as a function of size $N$. $N^\ddag$ is the location of the transition state for rearrangement and $F^\ddag$ is the barrier height. $\delta f$ represents the fluctuations caused by inhomogeneity of the driving force.}
\label{fig:FreeEnergy2}
\end{figure}

Much below the crossover temperature $T_A$, the interface between the rearranged and unrearranged region will become quite narrow and at the same time the surface energy cost will go up. Simultaneously, per unit volume the entropy gain achieved by rearranging a fixed number of particles goes down. This implies the droplet or instanton size will grow overall as the temperature is lowered below $T_A$. The free energy of the instanton involves a balance between a bulk term, corresponding to the rearranged core, and a surface term where steric conflicts remain between the rearranged core and the unrearranged surroundings. The size of the central region of the instanton becomes a decent reaction coordinate for describing an activated transition of escape from the original trapped state. If there are $N$ particles in the rearranging core region, then the free energy cost as a function of this reaction coordinate (shown in Figure~\ref{fig:FreeEnergy2}) becomes
\begin{equation}
    F(N) = -T S_c N + \gamma N^{2/3}.
\end{equation}

The driving force in the first bulk term involves the configurational entropy per particle $S_c$, reflecting a growing multiplicity of rearranged aperiodic states as the core grows in size. Particles in the interface have, however, have had to compromise between losing their vibrational entropy and stabilization energy or losing their configurational entropy. The surface term is called the “mismatch energy”. Xia and Wolynes showed how to estimate this mismatch free energy from density functional theory for molecular systems.\cite{2000Xia:2000kca} The magnitude they predicted depends on the vibrational freedom in the metastable state and turns out to agree well with simulations.\cite{biroli2008thermodynamic} We see the droplet argument implies that a rearranged region of size $N^\ddag= \left(\frac{2\gamma}{3T S_c} \right)^3$ will correspond with a thermodynamically accessible saddle point of the free energy and the free energy cost for reaching this saddle point grows as the configurational entropy density decreases upon cooling: $F^\ddag = \frac{4}{27}\frac{ \gamma^3}{(T S_c)^2}$.

On lowering temperatures, a larger and larger region must re-configure before irreversible configurational escape can be achieved. This growth in size causes the activation free energy barrier to grow quite rapidly upon cooling, so there is non-Arrhenius behavior of the rate. Qualitatively, these conclusions fit the phenomenology of how equilibrated supercooled liquids fall out of equilibrium to become glasses. Nevertheless, two key ingredients are missing from the simple instanton argument we just described. The first lacuna is that the heterogeneity of the starting aperiodic configurations changes the overall shape of the instanton from being a compact sphere. Another thing left out is that there will also be a dynamic interplay of instantonic droplets, which occurs on a longer time scale than the initial rearrangement. It is the latter interplay that smooths the dynamical transition at $T_A$.

\section{The Heterogeneous Aperiodic Crystal and the Random Field Magnet}
The overlap order parameter field corresponding to a self-generated random periodic structure will intrinsically be spatially inhomogeneous, even before any rearrangement has occurred. The plateau value $q$ should vary from place to place as the local packing structures, bonding patterns, etc. change spatially. Recent machine learning schemes try to describe such variations using terms like “softness”.\cite{Liu2019,tah2022fragility,cubuk2017structure} It has been shown that short molecular dynamics runs indeed do show variations in the local Debye-Waller factor throughout the sample. Owing to these variations in the initial values of order parameter, the corresponding gains in the configurational entropy from rearranging regions will also fluctuate as one scans across a locally metastable configuration. If the region is large enough, this variation can be estimated pretty accurately by finding the configurational heat capacity, which is the excess of the heat capacity over what one expects from the vibrations within a frozen metastable state. This quantity is accessible to measurement calorimetrically and indeed is used to determine the Kauzmann temperature. In addition, the mismatch energies themselves will fluctuate spatially, but this seems to be a smaller effect, at least for systems far from the high-pressure jamming point.

Since the free energy that would be released by reconfiguring a region varies randomly in space, unlike nucleation in a pristine system with a scalar order parameter, nucleation of new configurations somewhere in the glass initially resembles the heterogeneous nucleation of a bubble on a dirty surface with random quenched disorder. Escapes from metastable initial configurations resemble the activated transitions of  a magnet with random frozen fields acting on each site. The corresponding random field Ising magnet (RFM) is still metastable, because we are examining dynamics above the Kauzmann temperature, where there is a substantial free energy gain that comes from exploring more states. The spatial heterogeneity has several direct consequences, which we describe below. 

The instantons for a random field magnet are not precisely spherical. Instead, they vary in shape in different parts of liquid. In this way, they can take advantage of favorable fluctuations in the field. The interfaces between rearranged material and the surroundings, therefore, will broaden a bit to take advantage of the driving force fluctuations that have come from the self-generated randomness of the initial configuration. Owing to this broadening, the apparent surface tension is not a constant, but will decrease as the radius of curvature of the instanton droplet grows. Kirkpatrick, Thirumalai and Wolynes\cite{1989Kirkpatrick:1989iia} used the analogy to the RFIM to exploit the arguments and detailed calculations made by Villain for this surface accommodation effect.\cite{villain1985equilibrium} This ``wetting'' or broadening effect causes the mismatch energy term to vary more slowly with size, than the surface area. Instead, one finds from Villain’s argument that the mismatch energy cost scales in the same way with size as do the fluctuations in the driving force. The mismatch cost varies like $\gamma N^{1/2}$. Redoing the standard smooth instanton calculation for the free energy cost with this slower scaling with core size gives results both for the size of the rearranging regions and for the activation free energy for the random system that are different from those that start from a uniform $q$,
\begin{equation}
    N^\ddag= \left(\frac{\gamma}{2T S_c} \right)^2 \\ \text{ and }\\ F^\ddag = \frac{ \gamma^2}{T S_c}.
\end{equation}
There are many felicitous aspects about these two results. First, the correlation length or size of the droplets now grows like $S_c^{-2/d}$. A correlation length that scales in this way would restore consistency with hyperscaling, if we assume there is heat capacity discontinuity at the ideal glass transition. Such a discontinuity is both the result of the replica mean field calculations at the ideal glass transition and the extrapolated behavior at $T_K$, if we were to extrapolate the changes of the experimental heat capacities measured calorimetrically at the laboratory glass transition.

We also point out that numerical estimates of the length exponents for the random field Ising model are very close to the $\frac{2}{d}$ value predicted by the Villain scaling argument in three dimensions. 

The other felicitous result is that predicted barrier scaling, as $T_K$ is approached, exactly coincides with the empirical Vogel-Fulcher-Tammann relation that has been used so often for fitting data for supercooled liquid viscosity, 
\begin{equation}
    \log \tau = \log \tau_0 + \frac{D T_K}{T-T_K}
\end{equation}
This barrier scaling was also predicted by Adam and Gibbs. The random instanton argument, however, is quite distinct from the arguments made earlier by them.\cite{adam1965temperature} Adam and Gibbs made the reasonable, but somewhat arbitrary assumption that a reconfiguring region must have available to it, some fixed minimum threshold number of configurations, before it can rearrange, most naturally 2, one supposes. This assumption implies the rearranging region grows upon cooling, but the predicted size would still actually be quite small at the laboratory glass transition, a few molecules, perhaps. In contrast, the random instanton argument shows the reconfiguring region will have available to it a larger and larger number of configurations as the Kauzmann temperature $T_K$ is approached. This increase in the complexity of the rearranging region is consistent with inferences of the growth of reconfiguration volumes that have been obtained using bounds based on experimentally measured nonlinear responses.\cite{berthier2005direct} The inferred reconfiguration volumes from the RFOT argument are much larger than the very small ones envisioned by Adam and Gibbs because the complexity of the rearrangement increases upon cooling. The sizes predicted by the RFOT theory agree quite well both with direct observations of rearranging regions using microscopy at the surface of glasses\cite{ashtekar2010direct} and inferences of the sizes from local relaxation measurements by Israeloff\cite{vidal2000direct} and a host of fluctuation spectroscopies in bulk. 

Perhaps more important than this modification of the scaling relation, the heterogeneity of the initial aperiodic configuration leads to a very much more dramatic qualitative effect on the dynamics of glasses: the free energy cost of an instanton must vary in different parts of the system, which then will relax at different rates. This is called dynamic heterogeneity.  The variations of instanton cost throughout space provide the most parsimonious explanation of the well-known non-exponential character of the relaxation processes in glasses and supercooled liquids, which are considered by many to be the principal hallmark of glassy physics.

RFOT theory also allows one to estimate quantitatively the magnitude of these fluctuations of the rearrangement activation barriers. If we assume that the fluctuations are relatively modest, the resulting fluctuations in the barrier size can then be estimated to first order just by using the fluctuations of configurational entropy driving force for regions having the typical size $N^\ddag$. Of course, there well may be long tails to the barrier distribution. Since the typical barrier varies inversely to the configurational entropy density, $S_c$, one finds the typical fluctuations in the barrier depend on the configurational heat capacity:
\begin{equation}
    \frac{\delta F^\ddag}{F^\ddag} = \frac{\delta S_c}{S_c} = \frac{1}{S_c} \sqrt{\frac{k_{\mathrm{B}}\Delta C_p}{N^\ddag}}.
\end{equation}
As a consequence, substances with the large $\Delta C_p$ characteristic of ``fragile liquid'' have both very non-Arrhenius kinetics and very strongly nonexponential relaxations, while if $\Delta C_p$ is small (as it is for so called, ``strong liquids''), the relaxation time distribution will be narrower and the relaxation rates have nearly Arrhenius temperature dependence. By using microscopic estimates of the mismatch free energy costs Xia and Wolynes went further to deduce a simple quantitative relation between the exponent $\beta$ in the relaxation function $\exp[-(t/\tau)^\beta]$ and the rate at which the average barrier grows on cooling through the coefficient $D$ in the dimensionless form of the Vogel-Fulcher law Eqn. 5, which reflects the degree of non-Arrhenius behavior.\cite{xia2001microscopic} The predicted quantitative relaxation is reasonably well satisfied by the experimental data, although there is certainly room for fluctuations in the mismatch energies to play some role for specific substances. 

The effects of inhomogeneity that we have just described have been more formally addressed by using an explicit mapping of the free energy functional of a molecular fluid, onto the random field energy Ising magnet with an additional average field.\cite{stevenson2008constructing}  This mapping starts from Monasson’s two real replica approach to one step replica symmetry breaking.\cite{monasson1995structural}  Using Monasson’s starting point, Stevenson, Walczak, Hall and Wolynes mapped the density functional for an aperiodic minimum to a random Ising magnet, where the analog spins live on a lattice, which is itself random. The spins are located on the aperiodic lattice of the initial atomic configurations. After the mapping was made, they were able to take over RFIM results that have been obtained, in numerous other ways, for other, spins living on simpler lattices. In this explicit mapping, heterogeneous magnetic couplings were also generated, as well, and their magnitude could be estimated. Real space renormalization group calculation on the analog system, with both randomness in the interactions and in the fields, based on the works of Berker, show the randomness in the fields is the dominant contributor, at least for molecular fluids. This random field Ising analogy has also been developed further by Biroli, Tarjus and their coworkers using field theoretic approaches.\cite{biroli2018random,biroli2018randomb}

The effects of heterogeneity on activated events have also been studied analytically in a different way, using the two-replica tool by Dzero, Schmalian and Wolynes.\cite{2005Dzero:tp,2009Dzero:tr} The average instanton they computed, does not completely reproduce the results based on the random field magnet analogy. Instead of getting the fully renormalized surface tension, giving the $N^{1/2}$ mismatch scaling (which ultimately came from a renormalization group argument), this approach yields only a finite renormalization of the mismatch cost interface. It still yields a surface tension, but one smaller than the naive instanton result. Interestingly, the interface region of the fields theoretic replica instanton shows continuous replica symmetry breaking reminiscent of the Gardner transition. It is very hard to settle definitively the scaling of the mismatch energy with rearranging region size, since in real glasses and current simulations these regions never grow to be more than five molecular diameters. In the magnet analogy, this would allow only, at best, two “rounds” of real space renormalization, thus making the distinction between $N^{1/2}$ and $N^{2/3}$ scaling very delicate. The replica instanton calculations also allow computing the fluctuations in barrier height and gives results very much consistent with the previous argument, showing nonexponentiality depends on the configurational heat capacity.

\section{Strings or Filamentous Instantons}
Despite the spinodal being a crossover, understanding the dynamics of glasses at this border between activated dynamics and mode coupling theories is simultaneously interesting, conceptually important, and presently, still incomplete.\cite{2006Stevenson:2006kfa} This part of the phase diagram is the easiest part to access via computer simulation at reasonable computational cost, so it is important to understand this regime if we are to make proper inferences from such simulations about the asymptotic behavior on the longer time scales characteristic of laboratory glasses. Molecular systems enter this crossover region when their relaxation times have reached about a microsecond at normal temperatures. Colloidal glasses, on the other hand, are often prepared in this regime in the laboratory, again on, say, the one-hour time scale: the slowness of the motion of the individual colloid particles immersed in solvent, in comparison to small molecules, means they start out $10^6$ times slower at the microscopic scale. This regime is also relevant for biological soft matter, such as the cytoskeleton,\cite{2013Wang:vi} again, because of the large size of the constituents. While the spinodal crossover between activated and collisional motions in glasses can be approached by tuning temperature alone, it can also be accessed by imposing external mechanical forces. This is the origin of shear bands within RFOT theory.\cite{2017Wisitsorasak:wf}  The spinodal stress tuned spinodal explains the immense strength of glasses\cite{wisitsorasak2012strength} and many aspects of the unusual rheology and fluid mechanics of glassy systems.

The magnetic analogy we have discussed, allows us to study the crossover by exploiting the droplet theory of metastable magnets due to Michael Fisher,\cite{fisher1967theory} who used it to approximate critical exponents, and more explicitly, its extension to the spinodal by Klein and his collaborators.\cite{unger1984nucleation} This theory explicitly accounts for the complex shapes of droplets as the spinodal is approached where they become ``stringy'', as has been seen in computer simulations.\cite{2006Stevenson:2006kfa} 

The droplet theory in its present form is formulated at temperatures below the mean field spinodal. At the lowest temperatures, the core of the rearranging droplet is large, and its interface is relatively thin. But the core shrinks when the temperature is raised, and the interfacial region, at the same time, becomes wider. The rearranging region, therefore, becomes ever more filamentous as the crossover is approached from below. Like the compact droplet case, the free energy cost of rearranging a filamentous region has several contributions. Just as for the compact core there is a driving force proportional to the number of rearranging units, $N$, which depends on the bulk configurational entropy.

This part of the free energy of rearranging a region is the same as for the compact droplets remains $-T S_c N$. Because of the large surface of a filamentous object, however, the mismatch free energy cost scales also like N, rather than with a fractional power $N^{1/2}$ and $N^{2/3}$, as RFOT predicts for the compact clusters. The coefficient of this mismatch energy term estimated from the density functional arguments also will be somewhat larger than for the smoother interfaces of the more compact droplets, $F = \gamma' N$ with $\gamma' > \gamma$. Because $\gamma' > \gamma$, of course, at lower temperatures the compact rearranging regions will dominate.

The filamentous shapes are very much numerous, favoring them entropically at high temperatures over compact clusters. The shape entropies of either one-dimensional strings or percolation clusters, both of which approximate the lattice animals of the droplet theory, scale linearly with their size. Thus, if we consider only filamentous rearranging regions, the free energy profile,   
\begin{equation}
    F(N) = -T S_c N + \gamma' N -  T S_{\mathrm{shape}} N
\end{equation}
becomes linear in $N$: the average free energy profile for filamentous regions then either monotonically increases with $N$ or monotonically decreases with $N$ (if we leave out fluctuations of the driving force). The border between these two regions determines the spinodal crossover temperature, $T_c$, which, thus, satisfies the condition $T_c (S_c + S_{\mathrm{shape}}) = \gamma'$. 

Using microscopic estimates of $S_{\mathrm{shape}}$ and $\gamma'$, Stevenson, Schmalian and Wolynes show that this criterion gives a good estimate of the crossover temperatures of molecular fluids that had been inferred directly from mode coupling fits to experimental data.\cite{2006Stevenson:2006kfa} The temperature gap between the crossover and glass transition temperatures is largest for “strong” liquids and is smaller for more fragile liquids. More or less uniformly, the crossover for molecular systems is predicted to occur when the relaxation times reach about $10^{-7}$ seconds, so clearly activated dynamics and mode coupling physics coexist in this regime, where true glassy behavior begins to take over. Interestingly, the crossover in this picture is mathematically equivalent to the Hagedorn transition in string theories of elementary particles.\cite{hagedorn1965statistical,stevenson2010universal}

Heterogeneity, which broadens the free energy barriers for compact rearranging regions also favors filamentous rearrangements. The stringy excitations give rise to a low free energy tail to the barrier distribution. Stevenson and Wolynes show that many features of the so-called ``secondary relaxations'' in molecular glasses can be accounted for by assuming that these subdominant relaxations arise from such stringy instantons.\cite{stevenson2010universal} Their typical activation energy does not grow as fast upon cooling as the main part of the relaxation time distribution, so these secondary relaxations are most easily discerned experimentally in the low temperature glassy state, where the main relaxation is beyond the measurement time window.

\section{Aging, Rejuvenation, and Flames in Glasses}
So far, we have actually been discussing the nature of the activated events and relaxation processes in equilibrated systems, but glasses are not at equilibrium, but are kinetically frozen. As such, glasses still change after they have been initially prepared. They are said to ``age''.  In strict mean field theory, aging must be ascribed to some sort of hierarchical structure of the replica symmetry breaking. This is because the mean field barriers between states are infinite in the thermodynamic limit, so a glassy system will be completely trapped as soon as the replica symmetry is first broken; for further change to occur, new minima must somehow appear, giving the hierarchy. This does occur in the Gardner transition scenario.\cite{1987Kirkpatrick:1987cb} A simpler scenario, by Lubchenko and Wolynes, provides a good account of what is observed in glassy systems with finite range interactions.\cite{lubchenko2004theory} In their picture, the new minima are accessed not primarily by further global mean field replica symmetry breaking, but by having only some of the local regions rearrange to new nearly equilibrated parts of the landscape. The glass becomes a mosaic of aperiodic crystallites of different energies.

The most explicit way to develop the aging theory is to use the local library construction introduced by Lubchenko and Wolynes.\cite{lubchenko2004theory} A similar local energy landscape library view was also developed by Biroli and Bouchand.\cite{bouchaud2004adam} In its simplest form, this picture suggests that when the glass is first prepared, the glass is caught in a high energy state, one characteristic of equilibrium at the temperature at which it began to have trouble equilibrating. This state will be an aperiodic structure initially. The temperature at which the glass fell out of equilibrium is called the fictive temperature $T_f$. Upon further cooling then, the newly rearranged regions will dominantly be at the low temperature, but the starting state is no longer in equilibrium with the ambient temperature $T$. By being able to access lower temperature states, we see the core of either a compact droplet or filamentous string in the nonequilibrium glass acquires an additional driving force per particle $\Delta \epsilon \equiv \epsilon(T_f) - \epsilon(T)$. The free energy profile for compact rearranging regions thus becomes 
\begin{equation}
    F(N) = (-T S_c - \Delta \epsilon) N + \gamma N^{1/2}.
\end{equation}
Because this additional driving force remains nearly fixed as the system cools, once the glass has fallen out of equilibrium, the rearranging regions do not grow much in size as the system is further cooled and the activation free energy barrier only grows slightly. The enthalpy of activation, characterizing the additional slowing upon cooling, becomes nearly constant.

Since further aging in the glass is very nearly an Arrhenius process, thermal aging of a glass phenomenologically resembles an ordinary chemical reaction in some respects. Nevertheless, the total free energy barrier does not decrease from what it was at $T_g$, so aging is quite slow and occurs on time scales much longer than the glass preparation time. Likewise, the distribution of barrier heights does not change much from what it was at the fictive temperature, where the glass initially fell out of equilibrium, so the process is nonexponential. Both of these predicted features agree with experimental observations. The RFOT theory also predicts that the activation enthalpy in the frozen glass at temperature $T$ is nearly proportional to what the activation free energy was at $T_f$. This connection implies the relaxation times, at the lower temperature, grow, as a power law of the time scale of preparing the sample, which is sometimes called sometimes the “waiting time”.  

A similar sort of power law relation between waiting time and non-equilibrium relaxation in the glass also appears in mean field treatments of Ising spin glasses, but there it is interpreted as occurring through a distinct hierarchical landscape mechanism.

Because of the local nature of the individual activated transitions in the aging glass, the description of the glass by a global fictive temperature can only be an approximate one. After some aging, the glass will have acquired a new source of spatial heterogeneity, the variations of the local fictive temperatures of those regions that have already taken advantage of the chance to re-equilibrate. Of course, after sufficient time, these rearranged regions will overlap, and the approximation of a uniformly set fictive temperature will again be a good one.

The addition of a global fictive temperature or more completely a local fictive temperature field implies the properties of nonequilibrium glass can be manipulated in many ways and are quite complex. To parody Tolstoy, while all equilibrium supercooled liquids are alike, nonequilibrium glasses are each nonequilibrium in their own way, depending on their history. This complexity is most clearly seen in the phenomenon of “rejuvenation”. Upon heating an aged glass, the glass does not retrace the energetic history, versus ambient temperature, it had when it was cooled. Instead, the energy content of the glass first lags behind its cooling curve for a while, and then rapidly overshoots, as the aged regions of the glass that have been stabilized by rearranging when held at their lowest temperatures must wait to reach a still higher ambient temperature to re-equilibrate to the higher ambient temperature, than when they were first energetically trapped and were less stable. These features are recapitulated by the spatially inhomogeneous field theoretic version of the RFOT theory,\cite{bhattacharyya2008facilitation,bhattacharyya2010subquadratic,peter2009spatiotemporal} which is based on combining a description of the activated events with mode coupling theory to describe the interactions between instantons. 

The dynamics of the spatiotemporal inhomogeneities in the glassy state are intricate because they involve two features missing from strict mean field theory: energetic heterogeneity and facilitation. In this account, so far, we have focused on the heterogeneity effects, but locality implies instantons couple, so that there is also an effect called “facilitation”, which comes from instanton interaction effects. Facilitation arises in the local RFOT picture because once a particular region has rearranged, its neighboring regions now see a new environment, one more energetically suitable to the overturned region. This means the mismatch energy of the neighbors has changed and likely, decreased, allowing a faster activation of the neighboring regions. A completely controlled treatment of this effect would be quite intricate to achieve. One approach is to return to the mode coupling theory, but to add to the perturbative nonlinear relation for the memory kernel, a short circuit channel, where motion occurs by activation. This was done in a spatially averaged form by Bhattacharyya, Bagchi and Wolynes. The extended mode coupling theory (which we call inhomogeneous RFOT/MCT), never fully freezes (as does traditional MCT without the activated short circuit), but instead goes over to the activated behavior of droplet/instantons at low temperature. This approach may seem ad hoc, but is necessary because the instantons themselves are nonperturbative, just as in semiclassical quantum theory. The strict mode coupling theory was derived perturbatively. A similar issue arises in quantum chromodynamics.\cite{schafer1998instantons}

The spatially averaged BBW equation on long time scales becomes a self-consistent nonlinear integrodifferential equation for the mobility $\mu$ (which normally contains the full correlation function). Just as the static theory promoted the plateau value of the time delayed structure function $q$ to a spatially varying field, we can, on this longer time scale, promote the mobility to being a spatiotemporally varying quantity $\mu(\underline{r},t)$, along with allowing the fictive temperature to vary also in space and time. We can then construct the nonlinear mode coupling self-consistency equations for the mobility field $\mu(\underline{r},t)$. BBW, in their treatment, assumed that $\mu$ is space and time translation invariant, leading to the average mobility. To treat the spatiotemporal inhomogeneities, we can make additional approximation that while the mobility field varies in space and time, it does so slowly. This simplification results in a nonlinear partial differential equation:
\begin{equation}
    \frac{\partial \mu}{\partial t} = \mu \xi_1^2 \nabla^2 \mu + \mu \xi_2 (\nabla \mu)^2 + \mu (\mu - \bar{\mu})
\end{equation}
where $\xi_1$ and $\xi_2$ depend on the dynamical correlation length and $\bar{\mu}$ is the self-consistent result that would be obtained from the spatially averaged BBW theory. This input mobility, $\bar{\mu}$ arises primarily from the activated events and thus has a very strong dependence on both the ambient temperature $T$ and the fictive temperature $T_f$, which also varies in space and time $\bar{\mu}(T_f, T)$. As in the Lubchenko-Wolynes aging theory, the fictive temperature field locally equilibrates to the ambient temperature, through the mobility field, at a rate that depends on the local mobility
\begin{equation}
    \frac{\partial T_f}{\partial t} = - \mu (T_f -T).
\end{equation}

Physically, owing to the strong fictive temperature dependence of $\bar{\mu}$, the dynamics arising from the coupling between these two nonlinear equations resembles what happens in the theory of ordinary chemical flames, where the diffusion of heat and combustible materials couples to the production of heat, through an exothermic reaction, whose rate, in turn, in that case, depends strongly on the local temperature. Much as flames do not initiate by cooling, the effects of mode coupling facilitation effects on the cooling dynamics of the liquids into the glassy state are very mild. In contrast, on rejuvenation by heating, the autocatalytic feedback between $T_f$ and $\mu$, as in combustion, leads to propagating fronts emanating from the reconfiguring regions.\cite{peter2009spatiotemporal} This front propagation phenomenon was discovered by Ediger in his experiments on the calorimetry of ultrastable glasses. Ultrastable glasses can be prepared in the laboratory by vapor disposition.

\begin{figure}
\centerline{\includegraphics[width=6cm]{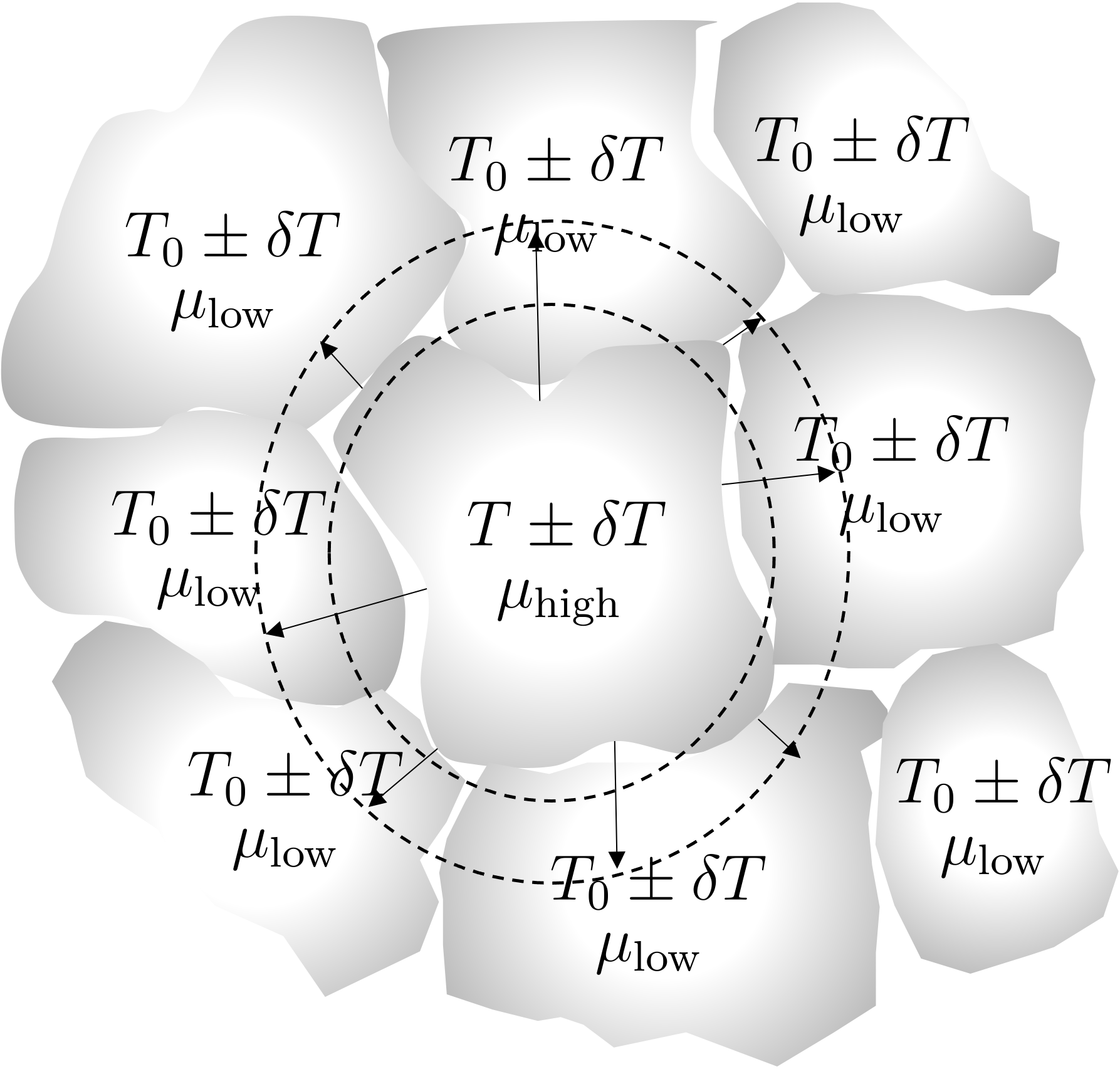}}
\caption{A cooperatively rearranging region once mobilized releases constraints on its neighbors, thus facilitating their rearrangement. Upon heating or rejuvenation of a glass, this leads to the propagation of mobility fronts. }
\label{fig:Tf}
\end{figure}

Vapor deposition leads to glasses with very low fictive temperatures, because glass dynamics near the free surface of the glass during the layered deposition process occurs much more rapidly than in the bulk glass would after it has been made.\cite{2013Wisitsorasak:wo} Upon heating such a glass, the surface again reconfigures more rapidly. The mobilized surface region upon heating, therefore, ignites a propagating front that penetrates the sample. For samples thicker than a few microns, this front will finally be intercepted by fronts that have started from reconfiguring regions that started in the bulk, but this leads to the surprisingly large length scale in the microns, not nanometers, which is the scale of the rearranging regions themselves. The velocity of the propagating mobility fronts quantitatively predicted by the inhomogeneous RFOT-MCT theory is roughly independent of the fictive temperature of the glass, but depends strongly on the relaxation rate in the mobilized region. This agrees with experiments!

\section{Glass Rigidity, Strength and Shear Bands}
On short time scales, a piece of glass is rigid, just as a periodic crystal is, which, in contrast, remains rigid, even on the longest time scales. As such, a glass object can still sustain a shear stress for a very long time. A modest degree of shear stress will distort the aperiodic lattice, and by compressing or stretching the local bonds, will increase the free energy of the glass. On the longer time scales, where thermally activated transitions will finally occur, the glass will flow eventually, like a liquid. These transitions will locally restore the aperiodic lattice to a non-stressed state (to the extent this is geometrically possible). Thus, like the stored energy in the glass that was preserved after quenching, the stress energy in the strained aperiodic lattice acts as an additional driving force for aging, hence, catalyzing the rearrangement and flow.\cite{wisitsorasak2012strength} On the longest times, then the nonequilibrium stressed glass will flow faster than a glass that has only been cooled, but not stressed.

The strain energy that is released when $N$ particles are allowed to relax to locally strain free aperiodic crystal scales with $N$
\begin{equation}
    \Delta E_{\mathrm{strain}} = \frac{\kappa \sigma^2}{2 G} N v.
\end{equation}
In this expression, $\sigma$ is the imposed shear stress, which is assumed to be uniform at large distance from the rearranging region, and $v$ is the volume of an atomic unit. According to elastic theory, owing to the long-range nature of elastic force interactions (which resemble elastic quadrupoles), the coefficient $\kappa$ depends on the shape of the region of $N$ particles. For compact spherical regions this shape dependent coefficient is given: $\kappa = 3 - \frac{6}{7-5 \nu}$ where $\nu$ is the Poisson ratio that determines the ratio of the bulk and shear moduli. While elastic interactions are long ranged, the shape dependence is still relatively weak, at most being logarithmically different between needles and spheres.

Both compact rearrangements and filamentous stringy rearrangements are catalyzed by imposed stress. The catalysis of stringy rearrangements implies that the spinodal threshold is shifted by the release of the imposed stress. RFOT theory, thus, puts a limit on the strength of glasses, when sufficient stress is applied to reach the spinodal.  Using this idea, Wisitsorasak and Wolynes have shown that at the ideal glass transition point, the strength of glass essentially matches the venerable estimate of maximum material strength of periodic materials, made by Frenkel a century ago. Ordinarily polycrystalline metals are much less strong then this Frenkel limit (by a factor of one hundred), owing to the existence of dislocations and grain boundaries. Indeed it was the discrepancy between Frenkel’s estimate and experimental measurements of strength that led to the idea of these defects.  The RFOT analysis shows, on the other hand, that ordinary glasses with a typical $T_g$ easily can achieve $1/3$ of the Frenkel maximum strength, in agreement with experiments. Glasses are, in fact, very strong, which is why they are used to make thinner (and cheaper!) objects than polycrystalline substances like ordinary metals. 

This large strength of glasses with respect to deformation means that when glasses are stressed to their limit, they may even break catastrophically by accessing new routes to dissipate stress. For example, with high stress, it is possible to create a small void (rather than the glass remaining densely packed), owing to the stored stress energy being sufficient to vaporize material. This is the reason we tend to think of the glass objects that we encounter around us as being breakable and easy to shatter.

If a sample of glass is able to remain intact and not shatter, but, nevertheless, nears the limiting stress, the glass can deform and flow, but it does so in an unusual inhomogeneous way: unlike an ordinary liquid, shear bands, where the displacements are concentrated, form.\cite{wisitsorasak2014dynamical,2017Wisitsorasak:wf} Wisitsorasak and Wolynes have described shear band formation by coupling visco elastic equations for the strain fields with the mobility transport and mobility production equations that we have just described as giving mobility fronts.

Since high stress catalyzes the production of mobility, the flame like propagation described previously leads to the spreading of mobility from the initially reconfiguring regions, which have been catalyzed by stress, producing a mobile band, where the displacements are concentrated. A shear band is mathematically a flame that has been ignited by stress. Failure by shear band formation is characteristic of metallic glasses. Electron microscopy of shear bands shows that they are quite sharp, and, in fact, vary on the nanometer length scale expected of reconfiguring regions. This, of course, suggests that for a fully quantitative description, one may have to go beyond the continuum style field theory treatment we have just described.

\section{Acknowledgments}
I would like to acknowledge the support for the work described in the chapter over many years from the NSF, most recently through the Center for Theoretical Biological Physics, sponsored by NSF grant PHY-2019745. The support from the Welch Foundation through the Bullard-Welch Chair at Rice University C-0016 is also acknowledged. I have enjoyed fruitful collaborations with many coworkers over the years, as can be seen in the references. I appreciate sharing this adventure with them. I would also single out Vas Lubchenko for many helpful discussions about glasses in recent years that particularly informed the framework of this chapter. I also thank Susan Merz for help with finalizing the MS and am grateful for the thoughtful preparation of figures by Dr. Andrei G. Gasic.

\bibliographystyle{ws-rv-van}
\bibliography{RSB40-PGW.bib}

\printindex                         

\end{document}